\documentclass[usegraphicx,usenatbib]{mn2e}

\newcommand{\aap}{A\&A}
\newcommand{\aj}{AJ}
\newcommand{\apj}{ApJ}
\newcommand{\apjl}{ApJ}
\newcommand{\apjs}{ApJS}
\newcommand{\araa}{ARA\&A}
\newcommand{\mnras}{MNRAS}
\newcommand{\nat}{Nat}
\newcommand{\physrep}{Phys. Rep.}

\newcommand{\prd}{Phys. Rev. D}

\title[The first  H~{\sc ii} regions]{The observational signature of the first H~{\sc ii} regions}

\author[Greif et al.]{Thomas H. Greif$^{1,4}$\thanks{E-mail: tgreif@ita.uni-heidelberg.de}, Jarrett L. Johnson$^{2,3}$, Ralf S. Klessen$^{1}$ and Volker Bromm$^{2,3}$\\$^{1}$ Zentrum f\"{u}r Astronomie der Universit\"{a}t Heidelberg, Institut f\"{u}r Theoretische Astrophysik, \\Albert-Ueberle-Stra\ss e 2, 69120 Heidelberg, Germany\\$^{2}$ Department of Astronomy, University of Texas, Austin, TX 78712, USA\\$^{3}$ Texas Cosmology Center, University of Texas, Austin, TX 78712, USA\\$^{4}$ Fellow of the International Max Planck Research School for Astronomy and Cosmic Physics at the University of Heidelberg}

\begin{document}

\maketitle
\topmargin-1cm

\begin{abstract}
We use three-dimensional smoothed particle hydrodynamics simulations together with a dynamical ray-tracing scheme to investigate the build-up of the first H~{\sc ii} regions around massive Population~III stars in minihaloes. We trace the highly anisotropic breakout of the ionizing radiation into the intergalactic medium, allowing us to predict the resulting recombination radiation with greatly increased realism. Our simulations, together with Press-Schechter type arguments, allow us to predict the Population~III contribution to the radio background at $\sim 100~{\rm MHz}$ via bremsstrahlung and 21-cm emission. We find a global bremsstrahlung signal of around $1~{\rm mK}$, and a combined 21-cm signature which is an order of magnitude larger. Both might be within reach of the planned Square Kilometer Array experiment, although detection of the free-free emission is only marginal. The imprint of the first stars on the cosmic radio background might provide us with one of the few diagnostics to test the otherwise elusive minihalo star formation site.
\end{abstract}


\section{Introduction}
One of the most important questions in modern cosmology is to understand how the first stars, the so-called Population~III (Pop~III), ended the cosmic dark ages at redshifts $z\la 30$ \citep{bl01,bl04a,cf05}. Their emergence led to a fundamental transformation in the early Universe, from its simple initial state to one of ever-increasing complexity. The emission from the hot, $T_{\rm eff}\sim 10^5~{\rm K}$, photospheres of Pop~III stars began the reionization of primordial hydrogen and helium in the intergalactic medium (IGM), although this process was completed only later on, when more massive galaxies formed \citep{fck06}. In addition, the supernova explosions that ended the lives of massive Pop~III stars distributed the first heavy elements into the IGM \citep{byh03,greif07,tfs07,wa08b}. This latter process might have had a significant impact on the physics of early star formation, as metal-enriched gas can cool more efficiently than primordial gas \citep{bl03a,omukai05,jappsen07a,jappsen09a}.

Based on numerical simulations, a general consensus has emerged that the first stars formed in dark matter minihaloes at $z\sim 20 - 30$, in isolation or at most as a small stellar multiple, and with typical masses of $M_{*}\sim 100~{\rm M}_{\odot}$ \citep[for a recent review, see][]{bromm09}. It is crucial to observationally test this key prediction. However, it has become evident that this will be very challenging. Even the exquisite near-IR ($\sim{\rm nJy}$) sensitivity of the upcoming {\it James Webb Space Telescope (JWST)} will not suffice to directly image such massive, single Pop~III stars \citep{bkl01,gardner06}, unless they explode as energetic pair-instability supernovae \citep{hw02,scannapieco05a}. The direct spectroscopic detection of recombination line emission from the H~{\sc ii} region surrounding the Pop~III star, as well as from the relic H~{\sc ii} region left behind once the star had died, is beyond the capability of {\it JWST} as well, although such line emission might be detectable from primordial stellar populations inside more massive host haloes \citep{schaerer02,schaerer03,johnson09}.

An alternative approach is to search for the global signature from many Pop~III stars that formed in minihaloes over large cosmic volumes \citep{hl97}. One such probe is the optical depth to Thomson scattering of cosmic microwave background (CMB) photons off free electrons along the line of sight, determined by the five-year {\it Wilkinson Microwave Anisotropy Probe (WMAP)} measurement to be $\tau_{e}\simeq 0.09\pm 0.02$ \citep{komatsu09}. This signal, however, is dominated by ionizing sources that must have formed closer to the end of reionization, with only a small contribution from Pop~III stars formed in minihaloes \citep{gb06,sbk08}. A second empirical signature is the combined bremsstrahlung emission from the H~{\sc ii} regions in their active and relic states around those minihaloes that hosted Pop~III stars. The resulting free-free radio emission leads to spectral distortions that might be detectable in the Rayleigh-Jeans part of the CMB spectrum. Recently, the ARCADE~2 experiment has attempted to measure such a free-free contribution from the epoch of the first stars \citep{kogut06}. The surprisingly strong signal found, however, cannot originate in early Pop~III stars, and in any case would overwhelm the much weaker contribution from the first stars and galaxies \citep{seiffert09}. The most promising detection strategy might be to scrutinize the background from the redshifted 21-cm line of neutral hydrogen \citep{fob06}. Once the central Pop~III star has died, the relic H~{\sc ii} region left behind would provide a bright source of 21-cm emission \citep{tokutani09}. Again, individual sources are much too weak to leave a detectable imprint, but the planned Square Kilometer Array (SKA) might be able to detect the cumulative signal \citep{furlanetto06,lazio08}.

We here carry out radiation hydrodynamics simulations of the evolution of H~{\sc ii} regions around massive Pop~III stars in minihaloes, giving us a detailed understanding of the properties of individual sources. We combine this with an approximate, Press-Schechter type analysis of the cosmological number density of minihaloes as a function of redshift to derive the observational signature of the first H~{\sc ii} regions as well as relic H~{\sc ii} regions, where we specifically focus on the free-free and 21-cm probes. We note that we do not include the feedback effects exerted by black holes or supernovae, which are possible end products of massive Pop~III stars \citep{hw02}. In this sense, we organize our work as follows. In Section~2, we describe the simulation setup and our implementation of the radiative transfer scheme in the smoothed particle hydrodynamics (SPH) code {\sc GADGET}-2 \citep{springel05}. In Section~3, we discuss the properties of the first H~{\sc ii} regions in their active as well as relic states and their observational signature in terms of recombination radiation, bremsstrahlung and 21-cm emission. Finally, in Section~4 we summarize our results and assess their implications. For consistency, all quoted distances are physical, unless noted otherwise.

\section{Numerical methodology}
Our treatment of ionizing and photodissociating radiation emitted by massive Pop~III stars is very similar to the methodology introduced in \citet{jgb07} and \citet{yoshida07}, with the exception that we here take the hydrodynamical response into account, self-consistently coupled to the chemical and thermal evolution of the gas. This allows us to model dense (D-type) as well as rarefied (R-type) ionization fronts, which is crucial for a proper treatment of the breakout of ionizing radiation. In the following, we describe our simulation setup, as well as the numerical implementation of the ray-tracing algorithm.

\subsection{Simulation set-up}
We perform our simulations in a cosmological box with linear size $200~{\rm kpc}$ (comoving), and $256^{3}$ particles per species, corresponding to a  particle mass of $\simeq 17~{\rm M}_{\odot}$ for dark matter and $\simeq 3~{\rm M}_{\odot}$ for gas. The simulations are initialized at $z=99$ with a fluctuation power spectrum determined by a $\Lambda$ cold dark matter ($\Lambda$CDM) cosmology with matter density $\Omega_{m}=1-\Omega_{\Lambda}=0.27$, baryon density $\Omega_{b}=0.046$, Hubble parameter $h=H_{0}/100~{\rm km}~{\rm s}^{-1}~{\rm Mpc}^{-1}=0.7$, where $H_{0}$ is the Hubble expansion rate today, and spectral index $n_{s}=0.96$ \citep{komatsu09}. We use an artificially high fluctuation power of $\sigma_{8}=1.6$ to accelerate structure formation in our relatively small box, although the cosmological mean is given by $\sigma_{8}=0.81$. We take the chemical evolution of the gas into account by following the abundances of H, H$^{+}$, H$^{-}$, H$_{2}$, H$_{2}^{+}$, He, He$^{+}$, He$^{++}$, and e$^{-}$, as well as the three deuterium species D, D$^{+}$, and HD.  We include all relevant cooling mechanisms, i.e. H and He collisional ionization, excitation and recombination cooling, bremsstrahlung, inverse Compton cooling, and collisional excitation cooling via H$_{2}$ and HD \citep{gj07}. We explicitly include H$_{2}$ cooling via collisions with protons and electrons, which is important for the chemical and thermal evolution of relic H~{\sc ii} region gas \citep{ga08}.

We run the simulations until the first minihalo in the box has collapsed to a density of $n_{\rm H}=10^{4}~{\rm cm}^{-3}$, at which point the gas has cooled to $\simeq 200~{\rm K}$ and becomes Jeans-unstable \citep{abn02,bcl02}. The first halo that fulfils this criterion collapses at $z_{*}\simeq 20$ and has a virial mass of $\simeq 9.4\times 10^{5}~{\rm M}_{\odot}$ and a virial radius of $\simeq 90~{\rm pc}$. Highly resolved simulations have shown that at later times, the gas condenses further under the influence of self-gravity to $n_{\rm H}\sim 10^{21}~{\rm cm}^{-3}$, where it becomes optically thick and forms a protostellar seed \citep{yoh08}. Due to its residual angular momentum, the central clump flattens and likely evolves into an accretion disk. In our case, we find a flattened structure already at a density of $n_{\rm H}=10^{4}~{\rm cm}^{-3}$ (see Fig.~1). Subsequently, the star grows to as massive as $\sim 100~{\rm M}_{\odot}$ within its lifetime of a few million years \citep{bl04b}. However, we note that under certain conditions the disk may fragment to form multiple objects of smaller masses \citep{cgk08}. Unfortunately, the details of the accretion phase and the concomitant radiative feedback are poorly understood, although some analytic investigations have been carried out \citep{tm04,mt08}. Under these circumstances, it seems best to initialize the calculation of the H~{\sc ii} region at the onset of the initial Jeans-instability, when the density exceeds $n_{\rm H}=10^{4}~{\rm cm}^{-3}$.

\begin{figure*}
\begin{center}
\resizebox{17cm}{8cm}
{\unitlength1cm
\begin{picture}(17,8)
\put(0.0,2.4){\includegraphics[width=5.6cm,height=5.6cm]{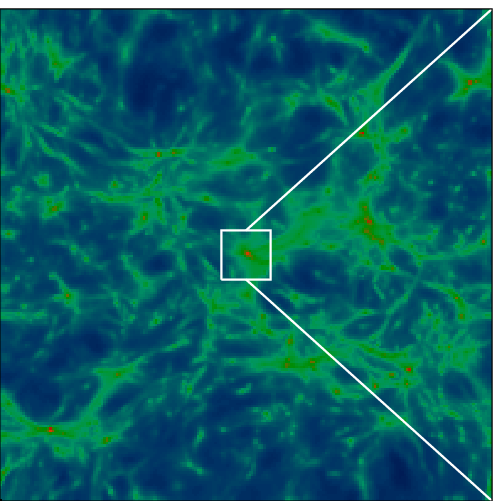}}
\put(5.7,2.4){\includegraphics[width=5.6cm,height=5.6cm]{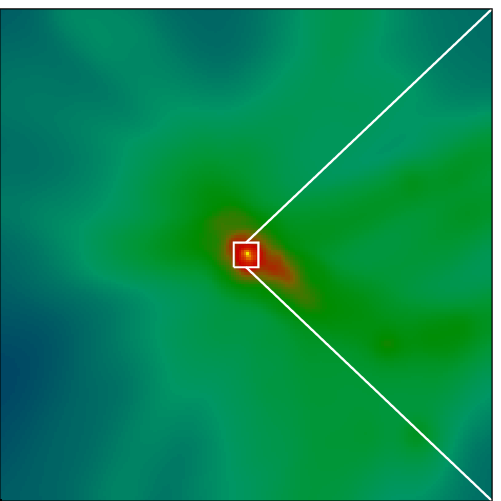}}
\put(11.4,2.4){\includegraphics[width=5.6cm,height=5.6cm]{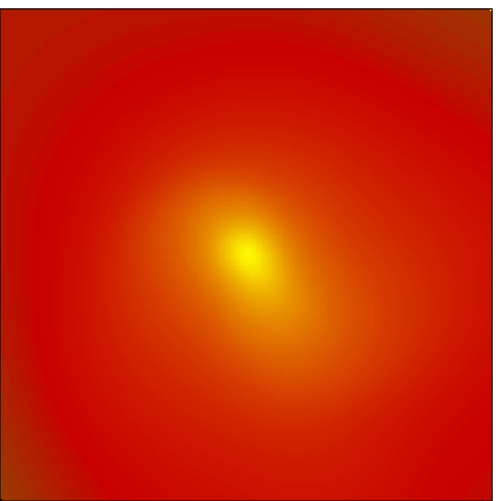}}
\put(0.0,0.0){\includegraphics[width=17cm,height=2.4cm]{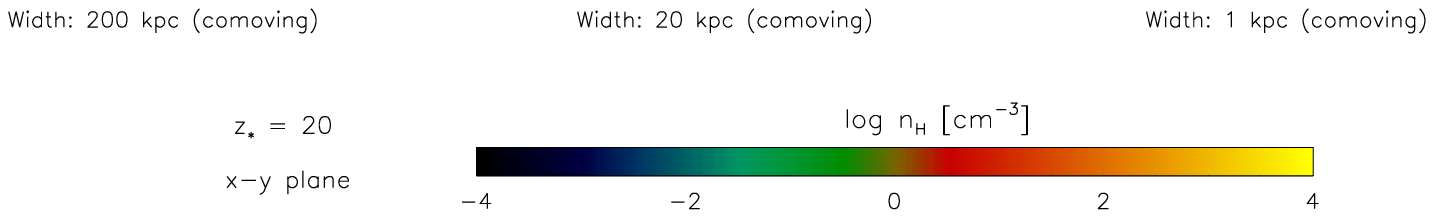}}
\end{picture}}
\caption{Sequential zoom-in on the first star-forming minihalo at $z_{*}\simeq 20$. Shown is the density-squared weighted average of the hydrogen density along the line of sight, just after the formation of the first Jeans-unstable clump in a $\simeq 9.4\times 10^{5}~{\rm M}_{\odot}$ minihalo. The flattening of the core due to angular momentum conservation during the collapse is marginally visible, with the consequence that ionizing radiation from the central source breaks out anisotropically (see Fig.~3).}
\end{center}
\end{figure*}

\subsection{Ray-tracing scheme}
The procedure used to calculate the Str\"{o}mgren sphere around the star for a given time-step $\Delta t$ is similar to the ray-tracing scheme used in \citet{jgb07}. We first designate an individual SPH particle as the source of ionizing radiation and create a spherical grid with typically $10^{5}$ rays and $500$ logarithmically spaced radial bins around the source particle. The minimum radius is set by the smoothing length of the central particle, while the maximum radius is chosen appropriately to encompass the entire H~{\sc ii} region. This approach may seem crude compared to existing methods that use adaptive grids \citep[e.g. {\sc HEALPIX};][]{gorski05}, but the increased angular and radial resolution towards the center tend to mirror the existing density profile. However, one must proceed with care if the ionization front encounters dense clumps far from the source, where the resolution may no longer be sufficient.

In a single, parallel loop, the Cartesian coordinates of all particles are converted to spherical coordinates, such that their density and chemical abundances are mapped to the bins corresponding to their radius, zenith angle and azimuth, denoted by $r$, $\theta$ and $\phi$, respectively. The volume of each particle is approximately given by $\Delta V\simeq h^{3}$, which transforms to $\Delta r=h$, $\Delta\theta=h/r$ and $\Delta\phi=h/(r\sin \theta)$. If the volume element of a particle intersects with the volume element of a bin, the particle contributes to the bin proportional to the density of the particle squared. This dependency ensures that overdense regions are not missed if the bin size is much larger than the smoothing length, which could occur far from the source where the grid resolution is poor. Accidental flash-ionization of minihaloes is thus avoided. Once the above steps are complete, it is straightforward to solve the ionization front equation along each ray:
\begin{equation}
n_{n}r_{\rm I}^{2}\frac{{\rm d}r_{\rm I}}{{\rm d}t}=\frac{\dot{N}_{\rm ion}}{4\pi}-\alpha_{\rm B}\int_{0}^{r_{\rm I}}n_{e}n_{+}r^{2}{\rm d}r\mbox{\ ,}
\end{equation}
where $r_{\rm I}$ denotes the position of the ionization front, $\dot{N}_{\rm ion}$ the number of ionizing photons emitted per second, $\alpha_{\rm B}$ the case B recombination coefficient, and $n_{n}$, $n_{e}$ and $n_{+}$ the number densities of neutral particles, electrons and positively charged ions, respectively. We assume that the recombination coefficient remains constant at its value for $10^{4}~{\rm K}$, which is roughly the temperature of the H~{\sc ii} and He~{\sc iii} region.

The numbers of H~{\sc i}/He~{\sc i} and He~{\sc ii} ionizing photons are given by
\begin{equation}
\dot{N}_{\rm ion}=\frac{\pi L_{*}}{\sigma T_{\rm eff}^{4}}\int_{\nu_{\rm min}}^{\infty}\frac{B_{\nu}}{h\nu}{\rm d}\nu\mbox{\ ,}
\end{equation}
where $h$ from now on denotes Planck's constant, $\sigma$ denotes Boltzmann's constant, and $\nu_{\rm min}$ is the minimum frequency corresponding to the ionization threshold of H~{\sc i} and He~{\sc ii}. We assume that massive Pop~III stars emit a blackbody spectrum $B_{\nu}$ (in ${\rm erg}~{\rm s}^{-1}~{\rm cm}^{-2}~{\rm Hz}^{-1}~{\rm sr}^{-1}$) with an effective temperature $T_{\rm eff}={\rm dex}(4.922$, $4.975$, $4.999)~{\rm K}$ and luminosity $L_{*}={\rm dex}(5.568$, $6.095$, $6.574)~{\rm L}_{\odot}$ for a $50$, $100$ and $200~{\rm M}_{\odot}$ star, respectively \citep{schaerer02}. This yields
\begin{equation}
\dot{N}_{\rm ion,HI/HeI}=[2.80,9.14,26.99]\times 10^{49}~{\rm s}^{-1}
\end{equation}
and
\begin{equation}
\dot{N}_{\rm ion,HeII}=[0.72,4.14,15.43]\times 10^{48}~{\rm s}^{-1}\mbox{\ .}
\end{equation}
We do not distinguish between the H~{\sc ii} and He~{\sc ii} region, which is a good approximation for massive Pop~III stars \citep{of06}. The lifetimes of the stars are given by $t_{*}=3.7$, $2.7$ and $2.2~{\rm Myr}$, respectively. We neglect the effects of stellar evolution, which might lead to a decrease of the number of ionizing photons emitted at the end of the main sequence \citep{marigo01,schaerer02}, although recent investigations have shown that rotating Pop~III stars remain on bluer evolutionary tracks and this effect might not be so strong \citep{yl05,wh06,vazquez07}.

To obtain a discretisation of the ionization front equation, we replace the integral on the right-hand side of equation~(1) by a discrete sum:
\begin{equation}
\int_{0}^{r_{\rm I}}n_{e}n_{+}r^{2}{\rm d}r\simeq\sum_{i}n_{e,i}n_{+,i}r_{i}^{2}\Delta r_{i}\mbox{\ ,}
\end{equation}
where $\Delta r_{i}$ is the radial extent of bin $i$, and the sum extends from the origin to the position of the ionization front at the end of the current time-step $\Delta t$. The above equation describes the advancement of the ionization front due to an excess of ionizing photons compared to recombinations. Similarly, the left-hand side of equation~(1), which models the propagation of the ionization front into neutral gas, is discretized by
\begin{equation}
n_{n}r_{\rm I}^{2}\frac{{\rm d}r_{\rm I}}{{\rm d}t}\simeq\frac{1}{\Delta t}\sum_{i}n_{n,i}r_{i}^{2}\Delta r_{i} \mbox{\ ,}
\end{equation}
where the sum now extends from the position of the ionization front at the previous time-step to its position at the end of the current time-step. We perform the above steps separately for the H~{\sc ii} and He~{\sc iii} region, since they require distinct heating and ionization rates. For the He~{\sc iii} region, we replace the quantities $n_{n}$ and $n_{+}$ in equation~(1) with $n_{n}=f_{\rm HeII}\,n_{\rm H}$ and $n_{+}=f_{\rm HeIII}\,n_{\rm H}$, where $f_{X}$ is the number density of species $X$ relative to $n_{\rm H}$. We adopt a case B recombination rate of $\alpha_{\rm B}=1.3\times 10^{-12}~{\rm cm}^{3}~{\rm s}^{-1}$ for He~{\sc iii} recombinations to He~{\sc ii} \citep{of06}. Applying the same prescription to the H~{\sc ii} region, we find $n_{n}=(f_{\rm HI}+f_{\rm HeI})\,n_{\rm H}$ and $n_{+}=(f_{\rm HII}+f_{\rm HeII})\,n_{\rm H}$. Similarly, we adopt a case B recombination rate of $\alpha_{\rm B}=2.6\times 10^{-13}~{\rm cm}^{3}~{\rm s}^{-1}$ for hydrogen and helium recombinations from their first ionized states to the ground state \citep{of06}. We initialize the calculation of the H~{\sc ii} region at the boundary of the He~{\sc iii} region, since hydrogen and helium are maintained in their first ionization states by recombinations of He~{\sc iii} to He~{\sc ii} \citep{of06}. We note that the exact position of the ionization front is not restricted to integer multiples of our pre-defined radial bins, but may instead lie anywhere in between. For this purpose we adopt a simple linear scaling of the number of ionizations and recombinations as a function of the relative position of the ionization front. The most expensive step in terms of computing time is the assignment of the density and the chemical abundances to the grid, while the ray-tracing itself requires only a negligible amount of time.

\subsection{Photoionization and photoheating}
Once the extent of the H~{\sc ii} and He~{\sc iii} region have been determined, the SPH particles within these regions are assigned an additional variable that stores their distance from the source. This information is then passed to the chemistry solver, which determines the ionization and heating rates, given by
\begin{equation}
k_{\rm ion}=\int_{\nu_{\rm min}}^{\infty}\frac{F_{\nu}\sigma_{\nu}}{h\nu}{\rm d}\nu
\end{equation}
and
\begin{equation}
\Gamma=n_{n}\int_{\nu_{\rm min}}^{\infty}F_{\nu}\sigma_{\nu}\left(1-\frac{\nu_{\rm min}}{\nu}\right){\rm d}\nu\mbox{\ ,}
\end{equation}
where $F_{\nu}$ and $\sigma_{\nu}$ denote the incoming specific flux and ionization cross section, respectively. For the case of a blackbody,
\begin{equation}
F_{\nu}=\frac{L_{*}}{4\sigma T_{\rm eff}^{4}r^{2}}B_{\nu}\mbox{\ ,}
\end{equation}
where $r$ is the distance from the source. The resulting rates are given by
\begin{equation}
k_{\rm ion,HI}=\frac{[0.45,1.32,3.69]\times 10^{-6}}{\left(r/{\rm pc}\right)^{2}}~{\rm s}^{-1}\mbox{\ ,}
\end{equation}
\begin{equation}
k_{\rm ion,HeI}=\frac{[0.42,1.43,4.29]\times 10^{-6}}{\left(r/{\rm pc}\right)^{2}}~{\rm s}^{-1}\mbox{\ ,}
\end{equation}
\begin{equation}
k_{\rm ion,HeII}=\frac{[0.67,3.72,13.57]\times 10^{-8}}{\left(r/{\rm pc}\right)^{2}}~{\rm s}^{-1}\mbox{\ ,}
\end{equation}
\begin{equation}
\Gamma_{\rm HI}=n_{\rm HI}\frac{[0.40,1.28,3.74]\times 10^{-17}}{\left(r/{\rm pc}\right)^{2}}~{\rm erg}~{\rm s}^{-1}~{\rm cm}^{-3}\mbox{\ ,}
\end{equation}
\begin{equation}
\Gamma_{\rm HeI}=n_{\rm HeI}\frac{[0.41,1.57,4.94]\times 10^{-17}}{\left(r/{\rm pc}\right)^{2}}~{\rm erg}~{\rm s}^{-1}~{\rm cm}^{-3}\mbox{\ ,}
\end{equation}
\begin{equation}
\Gamma_{\rm HeII}=n_{\rm HeII}\frac{[0.72,4.46,17.13]\times 10^{-19}}{\left(r/{\rm pc}\right)^{2}}~{\rm erg}~{\rm s}^{-1}~{\rm cm}^{-3}
\end{equation}
for a $50$, $100$ and $200~{\rm M}_{\odot}$ Pop~III star, respectively. These are taken into account every time-step, while the ray-tracing is performed only every fifth time-step. Since the hydrodynamic time-step is generally limited to one-twentieth of the sound-crossing time through the kernel, our treatment of the coupled evolution of the ionization front and the hydrodynamic shock is roughly correct. The computational cost of runs with and without ray-tracing are typically within a factor of a few.

\subsection{Photodissociation and photodetachment}
The final ingredient in our algorithm is the inclusion of molecule-dissociating radiation. This effect turns out to be of only minor importance in the present study, but will render our algorithm capable of addressing a general set of early Universe applications. Molecular hydrogen is the most important coolant in low-temperature, primordial gas, but is easily destroyed by radiation in the Lyman-Werner (LW) bands between $11.2$ and $13.6~{\rm eV}$. The small residual H$_{2}$ fraction in the IGM leads to a very small optical depth over cosmological distances, such that even a small background can have a significant effect \citep{har00,gb01,jgb07}. In our implementation, we do not take self-shielding into account, which becomes important for H$_{2}$ column densities $\ga 10^{14}~{\rm cm}^{-2}$ \citep{db96}. Such a high column density is difficult to achieve in minihaloes, and is more likely to occur within the virial radius of the first galaxies \citep{oh02}. However, the onset of turbulence in the first galaxies likely leads to a reduction of self-shielding via Doppler shifting \citep{wa07b,greif08}. For this reason we treat the photodissociation of H$_{2}$ in the optically thin limit, such that the dissociation rate in a volume limited by causality to a radius $r=c\,t_{*}$ is given by $k_{{\rm H}_{2}}=1.1\times 10^{8}F_{\rm LW}~{\rm s}^{-1}$, where $F_{\rm LW}$ is the integral of the specific flux $F_{\nu}$ over the LW bands, resulting in
\begin{equation}
k_{{\rm H}_{2}}=\frac{[1.27,3.38,9.07]\times 10^{-7}}{\left(r/{\rm pc}\right)^{2}}~{\rm s}^{-1}
\end{equation}
for a $50$, $100$ and $200~{\rm M}_{\odot}$ Pop~III star, respectively. We equate the photodissociation rate of hydrogen deuteride to that of molecular hydrogen. In the present work, we do not explicitly include photodetachment of H$^{-}$ and photodissociation of H$^{+}_{2}$, which might be problematic outside of the H~{\sc ii} region, where molecules survive collisional destruction. However, in the context discussed here, this caveat is not important \citep[see][]{jgb07}.

\section{Observational signature}
In the following, we discuss the direct observational signature of the first H~{\sc ii} regions and relic H~{\sc ii} regions in terms of recombination radiation, as well as their indirect signature in terms of a global radio background produced by bremsstrahlung and 21-cm emission.

\subsection{Build-up of H~{\sc ii} and He~{\sc iii} region}
The build-up of the first H~{\sc ii} regions by Pop~III stars in minihaloes was treated in one dimension by \citet{kitayama04} and \citet{wan04}, and in three dimensions by \citet{abs06}, \citet{awb07} and \citet{yoshida07}. The latter also treated the build-up of a smaller He~{\sc iii} region, which is created by the very hard spectrum of massive Pop~III stars. The consensus was that recombinations initially balanced ionizations within the virial radius of the host halo, leading to the formation of a D-type ionization front. Breakout occured after the density dropped sufficiently for the ionization front to race ahead of the hydrodynamic shock, becoming R-type. The hydrodynamic response of the gas is self-similar, since minihaloes approximately resemble singular isothermal spheres \citep{shu02,abs06}. The relevant parameters are set by the temperature of the singular isothermal sphere and the H~{\sc ii} region, which in our case are $T \simeq 200$ and $\simeq 10^{4}~{\rm K}$, respectively.

In Fig.~2, we compare the density profile of the \citet{shu02} solution to the simulation for the case of a $100~{\rm M}_{\odot}$ Pop~III star. Interestingly, we find a clear deviation from the ideal, spherically symmetric solution already during the D-type phase, which is caused by the anisotropic collapse of the minihalo. Due to angular momentum conservation, the gas spins up and forms a flattened, disk-like structure at a density of $10^{4}~{\rm cm}^{-3}$, which can be seen in the right panel of Fig.~1, and in the left panel of Fig.~2, where it is evident that the density dispersion is almost an order of magnitude within the central $\simeq 10~{\rm pc}$. In response to this anisotropy, which is further amplified by the density-squared dependence of recombinations, the ionization front first breaks out perpendicular to the disk, where the column density is lowest. This is visible in the left and middle panels of Fig.~3, as well as in Fig.~2, where the \citet{shu02} solution is approximately reproduced perpendicular to the disk, while the plane of the disk remains neutral and dense. Once the ionization front becomes R-type, spherical symmetry is asymptotically restored and the H~{\sc ii} region expands to $r_{\rm HII}\simeq 1.9$, $2.7$ and $3.7~{\rm kpc}$ for the $50$, $100$ and $200~{\rm M}_{\odot}$ Pop~III star, respectively. We find that He~{\sc ii} ionizing photons within the He~{\sc iii} region increase the central temperature by a factor of $\simeq 1.5$, leaving only a small imprint on the dynamical evolution of the H~{\sc ii} region \citep[see][]{yoshida07}. However, the He~{\sc ii} $\lambda 1640$ recombination line within the He~{\sc iii} region may be used as a distinct probe for the presence of massive Pop~III stars \citep{bkl01,oh01,tgs01,schaerer02}. In the following, we use the results obtained in this section to determine the recombination signature of the first H~{\sc ii} and He~{\sc iii} regions in their active as well as relic states.

 \begin{figure*}
 \begin{center}
 \resizebox{17cm}{5.67cm}
 {\unitlength1cm
 \begin{picture}(17,5.67)
 \put(0.0,0.0){\includegraphics[width=5.67cm,height=5.67cm]{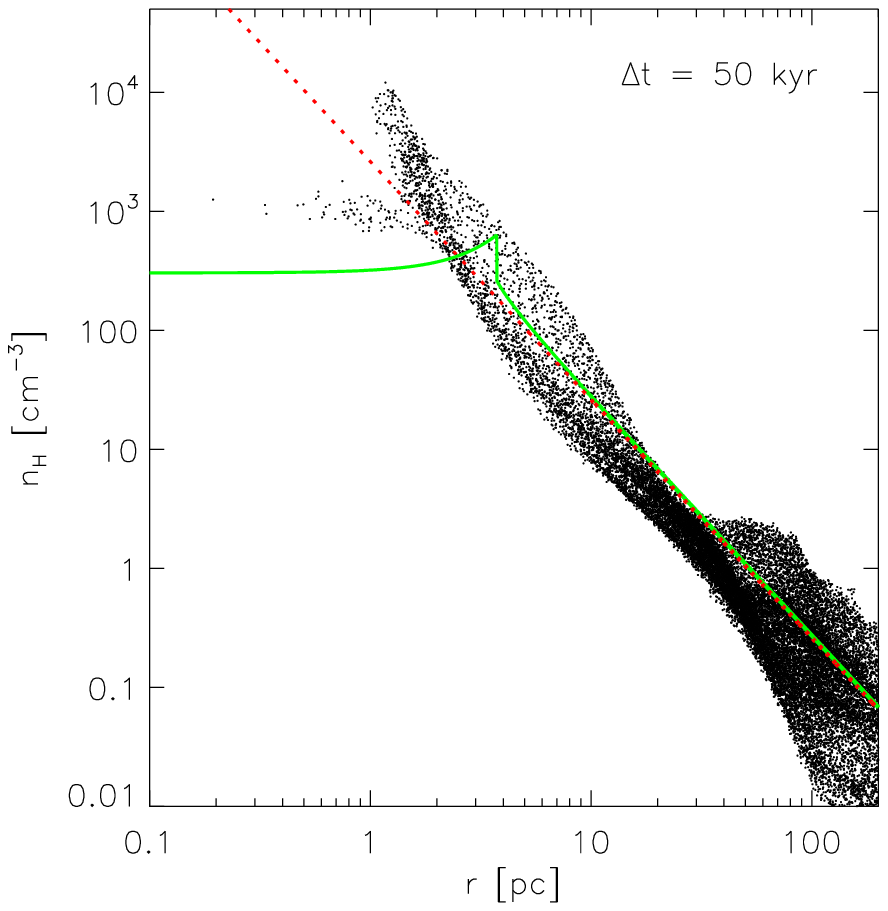}}
 \put(5.67,0.0){\includegraphics[width=5.67cm,height=5.67cm]{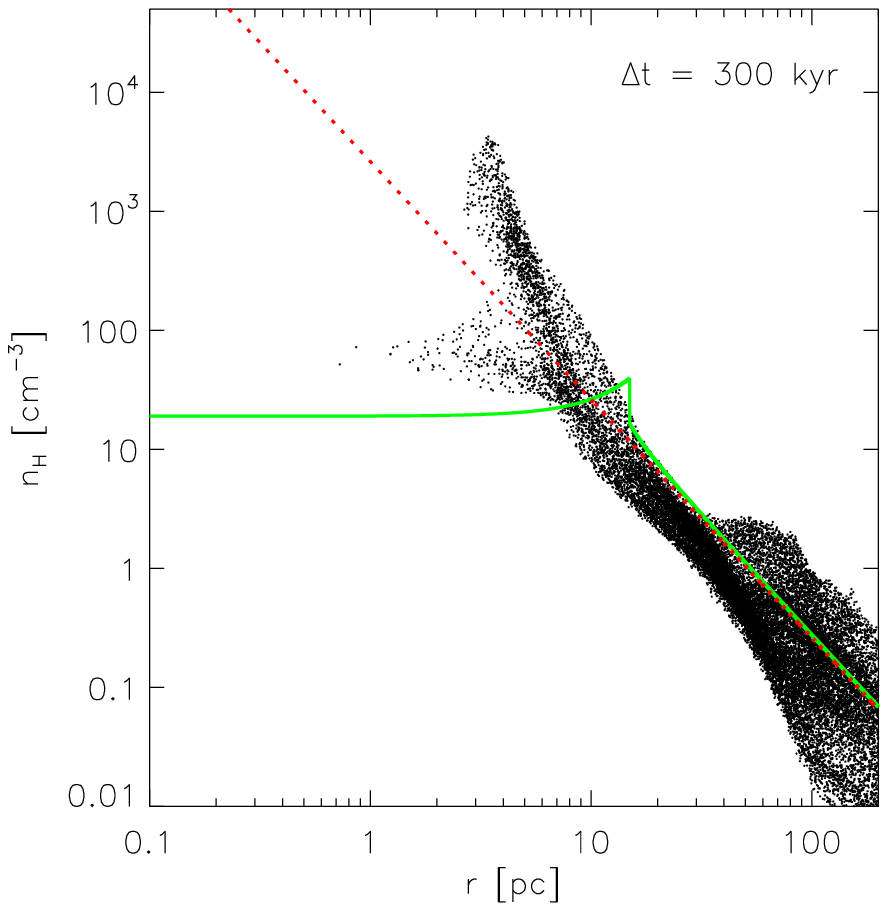}}
 \put(11.33,0.0){\includegraphics[width=5.67cm,height=5.67cm]{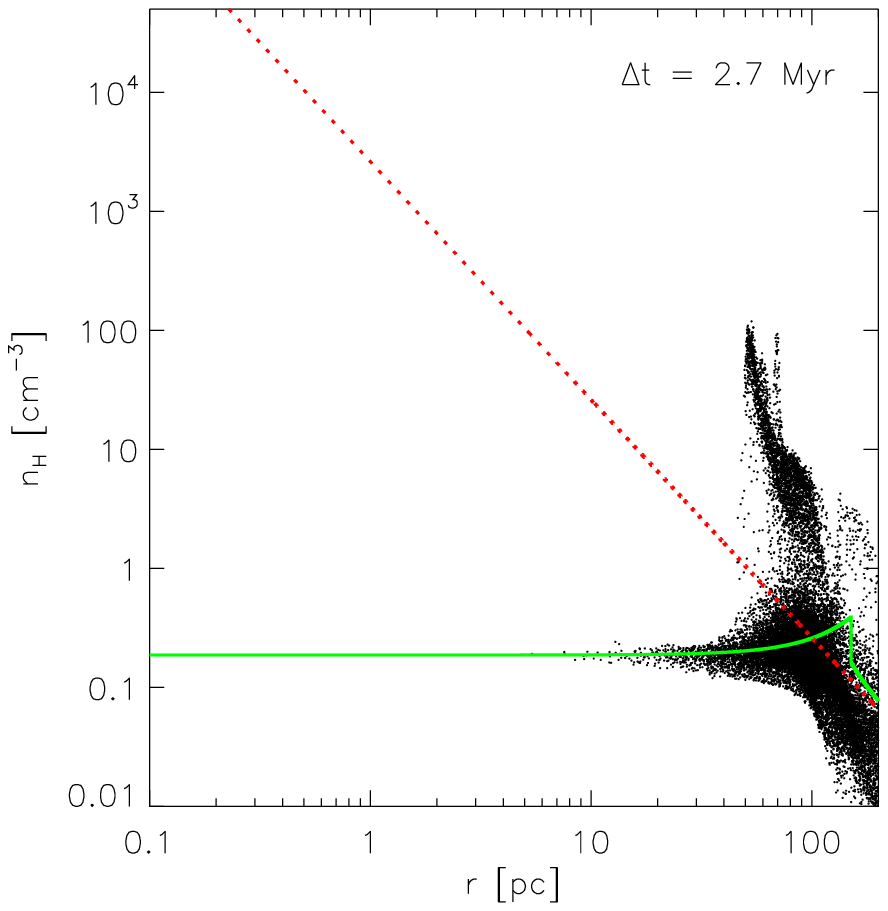}}
 \end{picture}}
 \caption{The hydrodynamic response of the gas to photoheating by a $100~{\rm M}_{\odot}$ Pop~III star after $50~{\rm kyr}$, $300~{\rm kyr}$, and $2.7~{\rm Myr}$ (from left to right). Shown is the hydrogen density as a function of radius for the simulation (black dots) and the analytic \citet{shu02} solution (green solid line), as well as the initial density profile of a singular isothermal sphere with $200~{\rm K}$ (red dotted line). The functional form of the analytic solution is reproduced perpendicular to the disk, where the ionization front breaks out after only a few $10~{\rm kyr}$. However, this is not the case in the plane of the disk, where the gas remains neutral and dense until the end of the star's lifetime.}
 \end{center}
 \end{figure*}

\begin{figure*}
\begin{center}
\resizebox{17cm}{20.2cm}
{\unitlength1cm
\begin{picture}(17,20.2)
\put(0.0,14.6){\includegraphics[width=5.6cm,height=5.6cm]{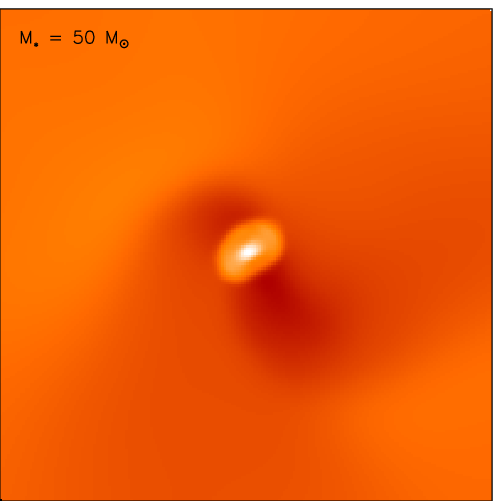}}
\put(5.7,14.6){\includegraphics[width=5.6cm,height=5.6cm]{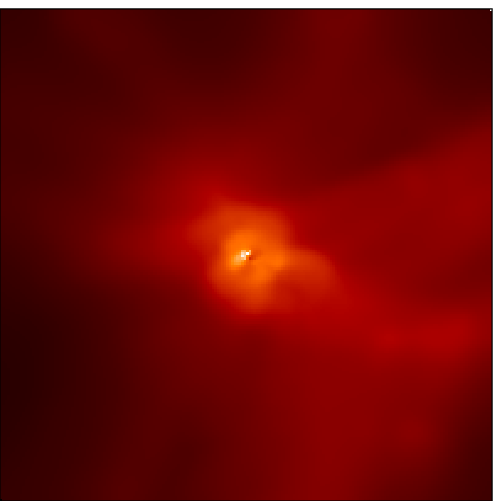}}
\put(11.4,14.6){\includegraphics[width=5.6cm,height=5.6cm]{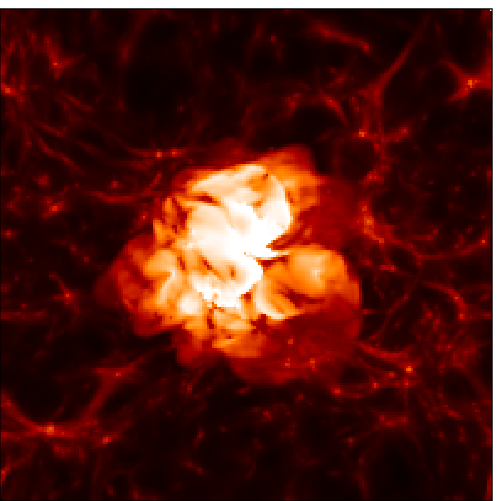}}
\put(0.0,8.8){\includegraphics[width=5.6cm,height=5.6cm]{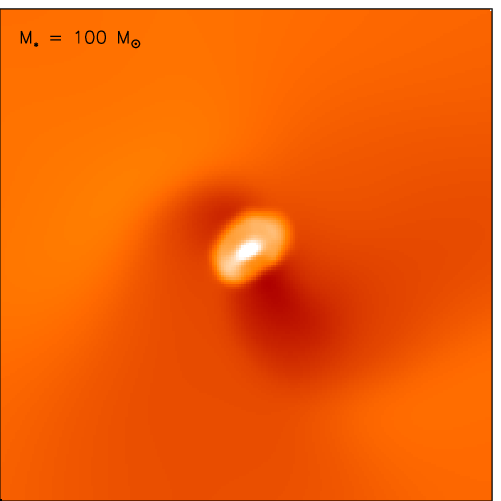}}
\put(5.7,8.8){\includegraphics[width=5.6cm,height=5.6cm]{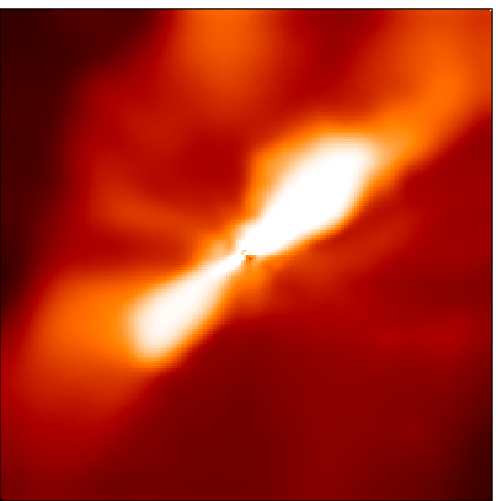}}
\put(11.4,8.8){\includegraphics[width=5.6cm,height=5.6cm]{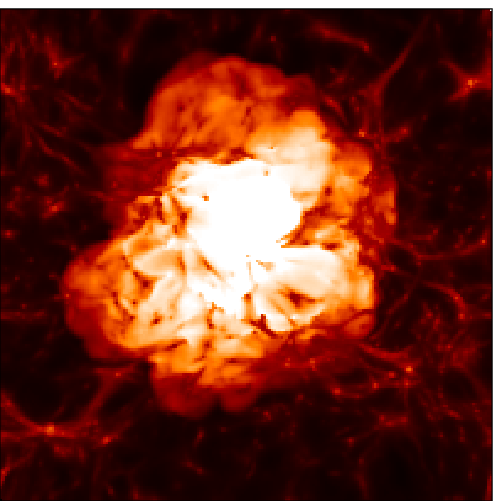}}
\put(0.0,3.0){\includegraphics[width=5.6cm,height=5.6cm]{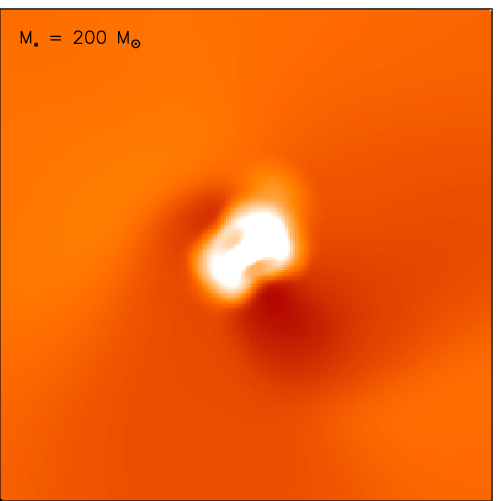}}
\put(5.7,3.0){\includegraphics[width=5.6cm,height=5.6cm]{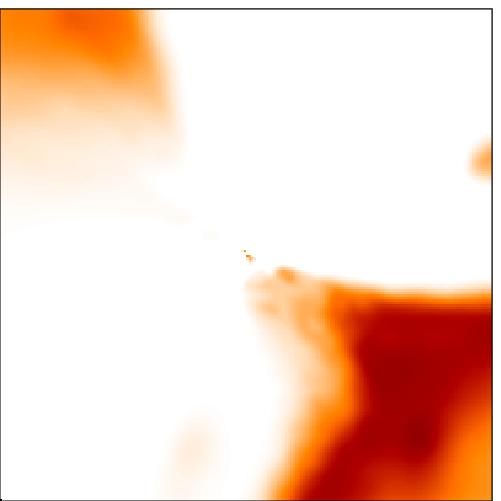}}
\put(11.4,3.0){\includegraphics[width=5.6cm,height=5.6cm]{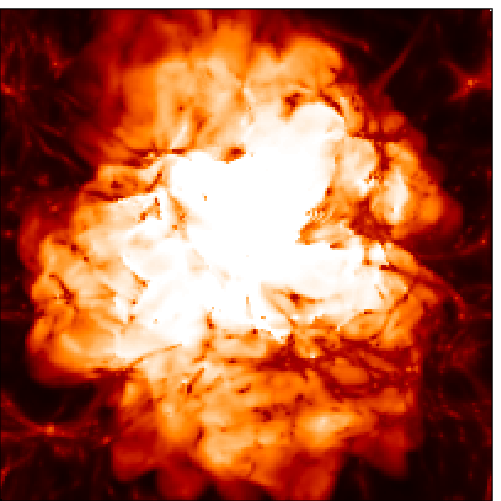}}
\put(0.0,0.0){\includegraphics[width=17cm,height=3cm]{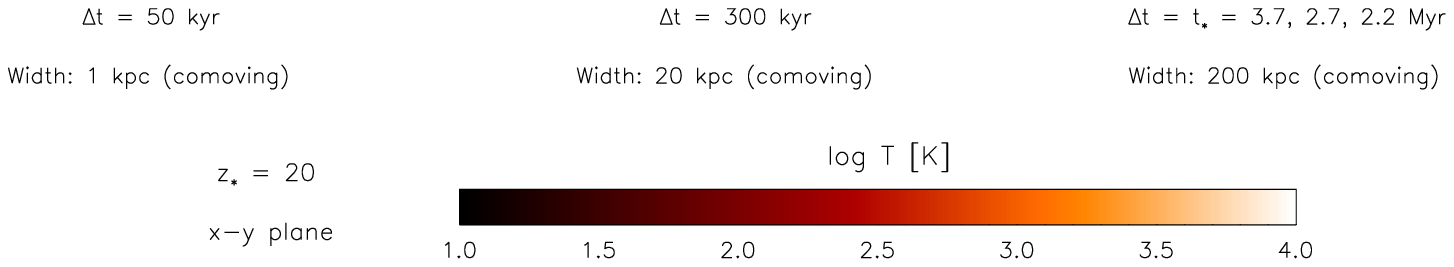}}
\end{picture}}
\caption{The H~{\sc ii} region created by a $50$, $100$ and $200~{\rm M}_{\odot}$ Pop~III star (from top to bottom) after $50~{\rm kyr}$, $300~{\rm kyr}$, and at the end of their lifetime (from left to right). Shown is the density-squared weighted average of the temperature along the line of sight. The spiral structure of the central clump as well as the resulting anisotropic breakout of the ionization front are clearly visible. For increasing stellar mass and ionizing photon output, breakout occurs earlier and is more isotropic. Once the ionization front becomes R-type, spherical symmetry is asymptotically restored and the H~{\sc ii} region expands to a final radius of $r_{\rm HII}\simeq 1.9$, $2.7$ and $3.7~{\rm kpc}$ after $3.7$, $2.7$ and $2.2~{\rm Myr}$, respectively.}
\end{center}
\end{figure*}

\subsection{Recombination radiation from individual H~{\sc ii} and He~{\sc iii} regions}
The strongest direct signature of the first H~{\sc ii} and He~{\sc iii} regions is likely generated by recombination radiation, since ionizing photons are absorbed by dense gas in the host halo. We here concentrate on the H$\alpha$ and He~{\sc ii} $\lambda 1640$ lines, since more energetic photons photons are scattered out of resonance by the neutral IGM, creating extended haloes around high-redshift sources \citep[e.g.][]{lr99}. The resulting fluxes may then be compared to the expected sensitivity of the Mid-Infrared Instrument (MIRI) on {\it JWST} at $\sim 10~\mu{\rm m}$ wavelengths \citep{gardner06}. The spatial resolution is limited by diffraction, such that a scale of $\simeq 1~{\rm kpc}$ at $z=20$ is marginally resolved, which allows us to approximate the region of emission as a point source. Using the simulation output, the total luminosities are given by
\begin{equation}
L_{{\rm H}\alpha}=j_{{\rm H}\alpha}\sum_{i}\frac{m_{i}}{\rho_{i}}\left(\frac{X\rho_{i}}{m_{\rm H}}\right)^{2}f_{e,i}f_{{\rm HII},i}
\end{equation}
and
\begin{equation}
L_{1640}=j_{1640}\sum_{i}\frac{m_{i}}{\rho_{i}}\left(\frac{X\rho_{i}}{m_{\rm H}}\right)^{2}f_{e,i}f_{{\rm HeIII},i}\mbox{\ ,}
\end{equation}
where $j_{{\rm H}\alpha}$ and $j_{1640}$ are the emissivity of the H$\alpha$ and He~{\sc ii} $\lambda 1640$ lines at $10^{4}~{\rm K}$ \citep{of06}, $X=0.76$ is the primordial mass fraction of hydrogen, $m_{\rm H}$ is the mass of the hydrogen atom, $m_{i}$ and $\rho_{i}$ are the mass and density of particle $i$, respectively, and the sum is over all particles in the simulation box. From the total luminosity, we determine the observed flux with the inverse-square law
\begin{equation}
F=\frac{L}{4\pi D_{\rm L}^{2}}\mbox{\ ,}
\end{equation}
where $D_{\rm L}$ is the cosmological luminosity distance. In Fig.~4, we show the observed flux for a $50$, $100$ and $200~{\rm M}_{\odot}$ Pop~III star as a function of time. The emission peaks before breakout, when the density in the host halo is still high, reaching a maximum flux of $\simeq 10^{-23}~{\rm erg}~{\rm s}^{-1}~{\rm cm}^{-2}$. Once the star turns off, the H$\alpha$ emission drops quite rapidly over the course of a few $10~{\rm Myr}$, while the He~{\sc ii} $\lambda 1640$ emission drops almost instantaneously in the $100$ and $200~{\rm M}_{\odot}$ cases, due to the high recombination coefficient of He~{\sc iii} to He~{\sc ii}. It is noteworthy that the emission in the He~{\sc ii} $\lambda 1640$ line is generally not much lower than that in the H$\alpha$ line, which may be used as an indicator for massive Pop~III stars \citep{bkl01,oh01,tgs01,schaerer02}.

For a $10\,\sigma$ detection with an exposure time of $100$ hours, the spectrograph on MIRI exhibits a typical limiting sensitivity of $\simeq 10^{-18}~{\rm erg}~{\rm s}^{-1}~{\rm cm}^{-2}$ \citep{panagia05}, implying that the first H~{\sc ii} regions are typically five orders of magnitude too faint for a direct detection. We must therefore resort to indirect methods that rely on their cumulative signal. One such signature is the cosmic infrared background (CIB), where the redshifted Ly$\alpha$ recombination photons from $z\sim 10$~--~$20$ might contribute at a detectable level \citep{sbk02,kashlinsky05}. Minihaloes, however, are not expected to be important sources for the CIB, as opposed to more massive dark matter haloes that host the first galaxies \citep{gb06}. This leads us to consider the radio background as a key diagnostic of the Pop~III minihalo formation site.

\subsection{Radio background produced by bremsstrahlung}
The first H~{\sc ii} regions in their active as well as relic states also emit bremsstrahlung via thermal motions of electrons in an ionized medium. In line with our conlusions of the previous section, the signature from an individual source is much too faint to be detected. However, the cumulative radio signal might be strong enough to be detected by the upcoming SKA. We will here further explore this possibility \citep[for a review of earlier work, see][]{fob06}.

Solving the cosmological radiative transfer equation, it is straightforward to derive a simple expression for the observed radio background $J_{\nu}$ (in ${\rm erg}~{\rm s}^{-1}~{\rm cm}^{-2}~{\rm Hz}^{-1}~{\rm sr}^{-1}$):
\begin{equation}
J_{\nu}=\int_{0}^{t_{H,0}}\frac{j_{\nu}}{(1+z)^{3}}c\,{\rm d}t\mbox{\ ,}
\end{equation}
where $t_{H,0}$ is the present Hubble time and $j_{\nu}$ is the specific emissivity of bremsstrahlung, given by
\begin{equation}
j_{\nu}=\epsilon_{\rm ff}\left<n_{e}^{2}\right>\left(T/10^{3}~{\rm K}\right)^{-1/2}\mbox{\ ,}
\end{equation}
where $\epsilon_{\rm ff}\simeq 10^{-39}~{\rm erg}~{\rm s}^{-1}~{\rm cm}^{3}~{\rm Hz}^{-1}~{\rm sr}^{-1}$, $\left<n_{e}^{2}\right>$ is the volume-averaged electron density, and $T$ is the temperature \citep{rl79}. We universally assume $T=10^{3}~{\rm K}$, since the relic H~{\sc ii} region cools quite rapidly to $\sim 10^{3}~{\rm K}$ via inverse Compton losses and adiabatic expansion once the star has died \citep[e.g.][]{greif07,yoshida07}. Furthermore, we assume $j_{\nu}=0$ at $z<6$, since photoheating during reionization evaporates minihaloes \citep{dijkstra04}. This leads to:
\begin{equation}
J_{\nu}=c\,\epsilon_{\rm ff}\int_{\infty}^{6}\frac{\left<n_{e}^{2}\right>}{\left(1+z\right)^{3}}\left|\frac{{\rm d}t}{{\rm d}z}\right|{\rm d}z\mbox{\ ,}
\end{equation}
where we relate $\left<n_{e}^{2}\right>$ to the number density of minihaloes according to:
\begin{equation}
\left<n_{e}^{2}\right> \simeq t_{\rm rec}\,n_{\rm H,b}^{2}\,V_{\rm HII}\left|\frac{{\rm d}N_{\rm ps}}{{\rm d}z}\right|\left|\frac{{\rm d}z}{{\rm d}t}\right|\mbox{\ .}
\end{equation}
Here, $t_{\rm rec}=\left(\alpha_{\rm B}\,n_{\rm H,b}\right)^{-1}$ denotes the recombination time for hydrogen atoms, $\alpha_{\rm B}$ the case B recombination rate for $T=10^{3}~{\rm K}$, $n_{\rm H,b}$ the background density, $N_{\rm ps}$ the number of minihaloes per comoving volume, $V_{\rm HII}=N_{\rm ion}/n_{{\rm H,b},0}$ the comoving volume of an individual H~{\sc ii} region in its active as well as relic state, which is independent of redshift, and $N_{\rm ion}=\dot{N}_{\rm ion}\,t_{*}$ the total number of ionizing photons emitted per Pop~III star (see Section~2). In the above equation, we have implicitly assumed that (relic) H~{\sc ii} regions survive for a recombination time, and that all ionizing photons escape into the IGM, which is a good approximation for massive Pop~III stars in minihaloes \citep{alvarez06}. We note that in the range of redshifts considered here, the recombination time is larger than the stellar lifetime and smaller than the age of the Universe. In principle, one must also account for the clustering of minihaloes (biasing), which reduces the net volume filling factor of H~{\sc ii} regions \citep{mw96,iliev03,gao05,reed05,gb06}. However, it is extremely difficult to determine the importance of this effect, since (i) the actual overlap depends on the relative separation of minihaloes, and (ii) previous ionization allows a nearby H~{\sc ii} region to become larger than usual. We therefore neglect biasing, but keep in mind that the actual signal may be somewhat lower.

\begin{figure}
\begin{center}
\includegraphics[width=8.5cm]{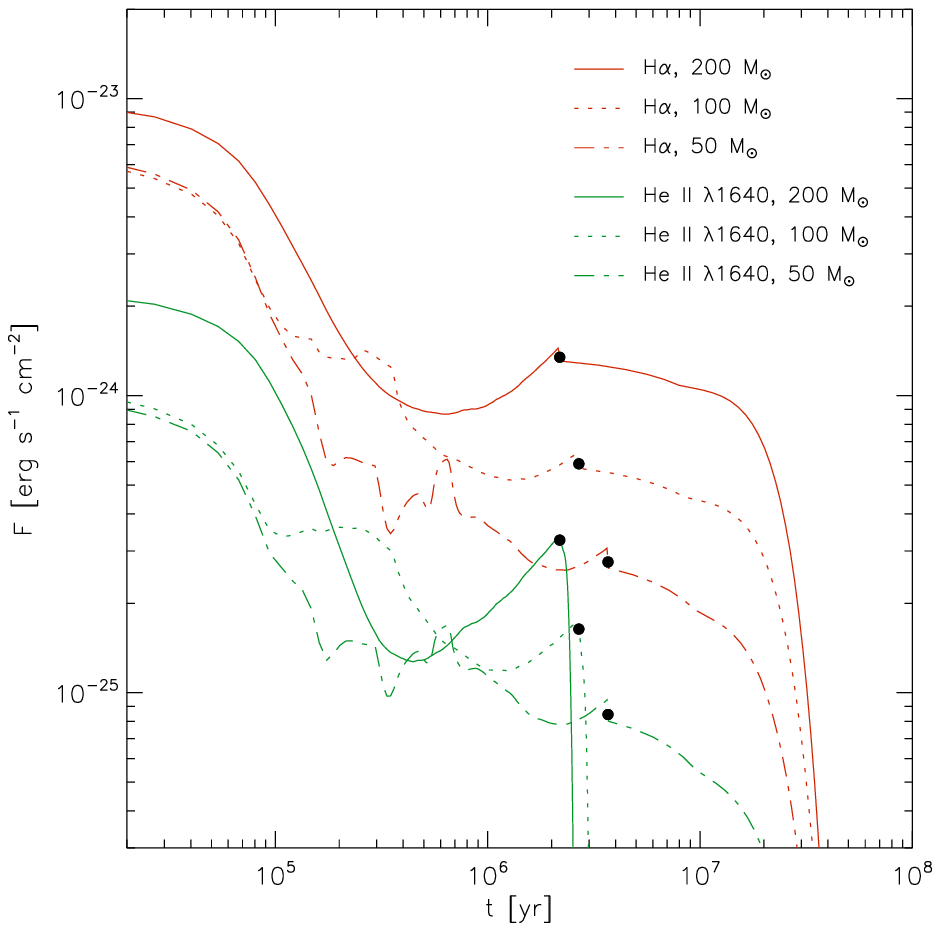}
\caption{The observed recombination flux in H$\alpha$ and He~{\sc ii} $\lambda 1640$ for a $50$, $100$ and $200~{\rm M}_{\odot}$ Pop~III star, shown as a function of time after reaching the zero-age main-sequence (the black dots denote the end of their lifetimes). The emission peaks before breakout, when the density in the host halo is still high, reaching a maximum flux of $\simeq 10^{-23}~{\rm erg}~{\rm s}^{-1}~{\rm cm}^{-2}$. Once the star turns off, the H$\alpha$ emission drops quite rapidly over the course of a few $10~{\rm Myr}$, while the He~{\sc ii} $\lambda 1640$ emission drops almost instantaneously in the $100$ and $200~{\rm M}_{\odot}$ cases, due to the high recombination coefficient of He~{\sc iii} to He~{\sc ii}. The emission in the He~{\sc ii} $\lambda 1640$ line is generally not much lower than that in the H$\alpha$ line, which is characteristic for a top-heavy IMF and may be used as an indicator for massive Pop~III stars. For a $10\,\sigma$ detection and an exposure time of $100$ hours, the limiting sensitivity of the MIRI spectrograph on {\it JWST} is approximately $10^{-18}~{\rm erg}~{\rm s}^{-1}~{\rm cm}^{-2}$, indicating that the first H~{\sc ii} regions are typically five orders of magnitude too faint to be detected directly.}
\end{center}
\end{figure}

In equation~(23), the number density of minihaloes is given by
\begin{equation}
N_{\rm ps}\left(z\right)=\int_{M_{\rm min}}^{M_{\rm max}}n_{\rm ps}\left(z,M\right){\rm d}M\mbox{\ ,}
\end{equation}
where $n_{\rm ps}$ is the well-known Press-Schechter mass function \citep{ps74}. The minimum mass required for efficient cooling within a Hubble time may be found in \citet{yoshida03a} and \citet{ts09}:
\begin{equation}
M_{\rm min}\simeq 10^{6}~{\rm M}_{\odot}\left(\frac{1+z}{10}\right)^{-2}\mbox{\ ,}
\end{equation}
while the maximum mass is set by the requirement that cooling must be dominated by molecular hydrogen, i.e. the virial temperature must not exceed $T \simeq 10^{4}~{\rm K}$ for atomic hydrogen cooling, resulting in \citep[e.g.][]{bl01}
\begin{equation}
M_{\rm max}\simeq 2.5\times 10^{7}~{\rm M}_{\odot}\left(\frac{1+z}{10}\right)^{-3/2}\mbox{\ .}
\end{equation}
We have found that our results are only marginally affected by the upper mass limit, but depend sensitively on the lower mass limit, since most minihaloes reside at the lower end of the halo distribution function.

After combining the above equations, we obtain
\begin{equation}
J_{\nu}\simeq \frac{c\,\epsilon_{\rm ff}\,N_{\rm ion}}{\alpha_{\rm B}}N_{\rm ps}\left(z=6\right)\mbox{\ ,}
\end{equation}
which, for an IMF consisting solely of $100~{\rm M}_{\odot}$ Pop~III stars, yields
\begin{equation}
J_{\nu}\simeq 300~{\rm mJy}~{\rm sr}^{-1}\mbox{\ .}
\end{equation}
The brightness temperature, $T_{b}=c^{2}J_{\nu}/2k_{\rm B}\nu^{2}$, is given by
\begin{equation}
T_{b}\simeq 1~{\rm mK}\left(\frac{\nu}{100~{\rm MHz}}\right)^{-2}\mbox{\ .}
\end{equation}
In the following, we investigate whether a signal of this magnitude is observable by the upcoming SKA.

The sensitivity of radio instruments is generally defined by the ratio of the effective collecting area $A_{e}$ to the system temperature $T_{\rm sys}$. For the SKA with its proposed aperture array configuration at low frequencies, $A_{e}/T_{\rm sys}\simeq 5\times 10^{3}~{\rm m}^{2}~{\rm K}^{-1}$ at $100~{\rm MHz}$ \footnote{http://www.skatelescope.org}. In this range, the system temperature is dominated by Galactic synchrotron emission, for which a useful approximation is given by $T_{\rm sky}\simeq 180~{\rm K}\left(\nu/180~{\rm MHz}\right)^{-2.6}$ \citep{fob06}, resulting in $T_{\rm sys}\simeq 800~{\rm K}$ and $A_{e}\simeq 4\times 10^{6}~{\rm m}^{2}$. The minimum angular resolution for an array filling factor of unity at $100~{\rm MHz}$ is approximately $15~{\rm arcmin}$. At higher resolutions, the sensitivity decreases much too rapidly for effective imaging. In Fig.~5, we compare the sensitivity of the SKA for a $10\,\sigma$ detection, a bandwidth of $\Delta\nu_{\rm obs}=1~{\rm MHz}$, and an integration time of $1000~{\rm h}$ to the brightness temperature and specific flux expected for free-free emission. Although the figure implies that the free-free signal is detectable by the SKA, we have neglected biasing as well as radiative feedback in the form of a global LW background, which attenuates star formation in minihaloes \citep{jgb07,jgb08}. Another complicating issue is the overlap with 21-cm emission, which makes it nearly impossible to isolate the contribution from bremsstrahlung. In consequence, we do not believe that this signal will be observable in the near future.

\begin{figure}
\begin{center}
\includegraphics[width=8.5cm]{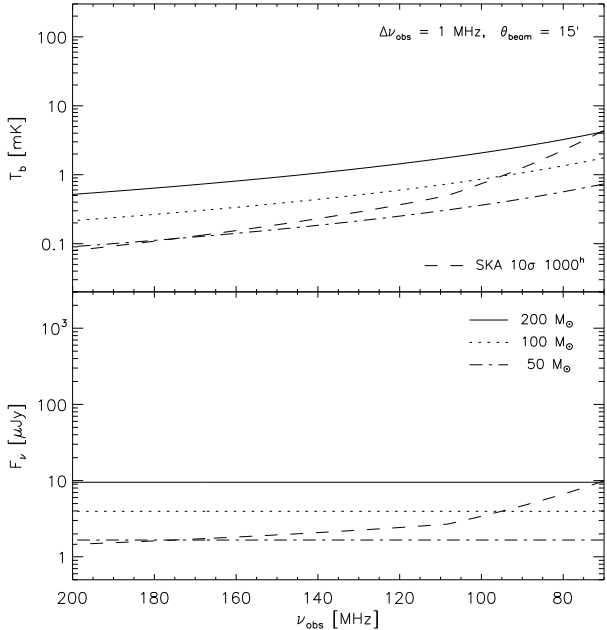}
\caption{The brightness temperature and specific flux of the radio background produced by bremsstrahlung, shown as a function of observed frequency. We have chosen a beam size of $15~{\rm arcmin}$ to achieve the highest possible resolution and sensitivity at $100~{\rm MHz}$ for the currently planned configuration of the SKA. The dot-dashed, dotted and solid lines correspond to an initial mass function consisting solely of $50$, $100$ and $200~{\rm M}_{\odot}$ Pop~III stars, respectively. The dashed line shows the sensitivity of the SKA for a $10\,\sigma$ detection, a bandwidth of $1~{\rm MHz}$, and an integration time of $1000~{\rm h}$. Although the free-free signal is in principle detectable by the SKA, we have here neglected biasing and radiative feedback, which act to reduce the signal. For this reason we do not believe that the free-free signal of the first H~{\sc ii} regions in their active or relic states will be observable in the near future.}
\end{center}
\end{figure}

\subsection{Radio background produced by 21-cm emission}
Perhaps the most promising observational signature comes from 21-cm emission of the relic H~{\sc ii} region gas once the star has died, a prospect that was already investigated by \citet{tokutani09}. An emission signal requires the spin temperature $T_{S}$ of neutral hydrogen to be greater than the temperature of the CMB, with its relative brightness determined by $T_{S}$ and the size of the relic H~{\sc ii} region. The spin temperature is set by collisional coupling with neutral hydrogen atoms, protons and electrons, as well as radiative coupling to the CMB. Furthermore, it may be modified by the so-called Wouthuysen-Field effect, which describes the mixing of spin states due to the absorption and re-emission of Ly$\alpha$ photons \citep{wouthuysen52,field59}. The color temperature of the Ly$\alpha$ background is determined by the ratio of excitations to de-excitations, which approaches the kinetic gas temperature at high redshifts, where the optical depth to Ly$\alpha$ scattering is very large \citep{fob06}. In this case, adopting the Rayleigh-Jeans approximation and assuming $T_{S}\gg T_{*}$, where $T_{*}=h\nu_{21}/k_{\rm B}=68~{\rm mK}$ is the temperature associated with the 21-cm transition, the spin temperature may be written as \citep{mmr97}
\begin{equation}
T_{S}=\frac{T_{\gamma}+\left(y_{c}+y_{\alpha}\right)\,T}{1+y_{c}+y_{\alpha}}\mbox{\ ,}
\end{equation}
where $T_{\gamma}$ is the temperature of the CMB. The collisional coupling coefficient $y_{c}$ is approximately given by
\begin{equation}
y_{c}=\frac{T_{*}}{A_{21}T}\left(n_{\rm HI}\kappa_{\rm HI}+n_{e}\kappa_{e}\right)\mbox{\ ,}
\end{equation}
where $A_{21}=2.85\times 10^{-15}~{\rm s}^{-1}$ is the Einstein A-coefficient for the 21-cm transition, and $\kappa_{\rm HI}$ and $\kappa_{e}$ are the effective single-atom rate coefficients for collisions with neutral hydrogen atoms and electrons, respectively. Good functional fits in the temperature range $100~{\rm K}\la T\la 10^{4}~{\rm K}$ are given by
\begin{equation}
\kappa_{\rm HI}=10^{-11}\,T^{1/2}~{\rm cm}^{3}~{\rm s}^{-1}
\end{equation}
and
\begin{equation}
\kappa_{e}=2\times 10^{-10}\,T^{1/2}~{\rm cm}^{3}~{\rm s}^{-1}\mbox{\ ,}
\end{equation}
which we have obtained from the rates quoted in \citet{kmm06}. At $z\la 20$, the electron fraction in the IGM remains above $f_{e}=0.1$ for most of the lifetime of the relic H~{\sc ii} region. In this case, the collisional coupling coefficient is given by
\begin{equation}
y_{c}\simeq 0.015\,\left(\frac{f_{e}}{0.5}\right)\,\left(\frac{T}{10^{3}~{\rm K}}\right)^{-1/2}\,\left(\frac{1+z}{10}\right)^{3}\mbox{\ .}
\end{equation}
A derivation of the Ly$\alpha$ coupling coefficient $y_{\alpha}$ requires radiative transfer of local as well as global Ly$\alpha$ radiation, which is beyond the scope of this work. We therefore consider two limiting cases: one in which we only consider collisional coupling, and the other in which a strong Ly$\alpha$ background drives the spin temperature towards the gas temperature (i.e. $y_{\alpha}\gg 1$ or $T_{S}=T$).

\begin{figure*}
\begin{center}
\resizebox{17cm}{13.9cm}
{\unitlength1cm
\begin{picture}(17,13.9)
\put(0.0,8.3){\includegraphics[width=5.6cm,height=5.6cm]{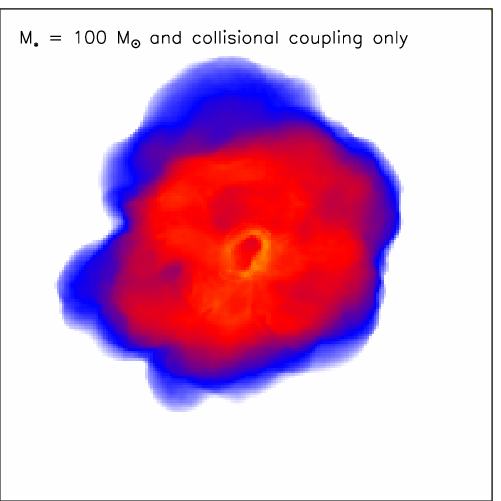}}
\put(5.7,8.3){\includegraphics[width=5.6cm,height=5.6cm]{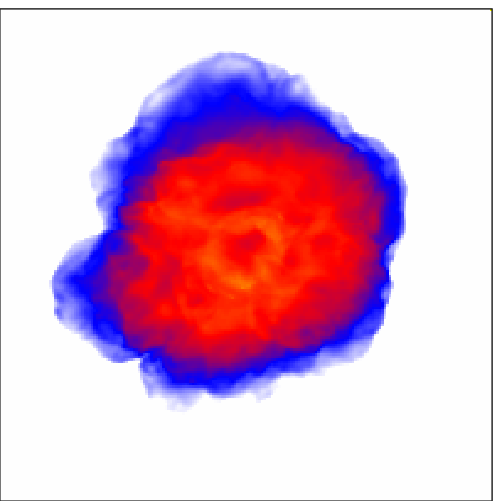}}
\put(11.4,8.3){\includegraphics[width=5.6cm,height=5.6cm]{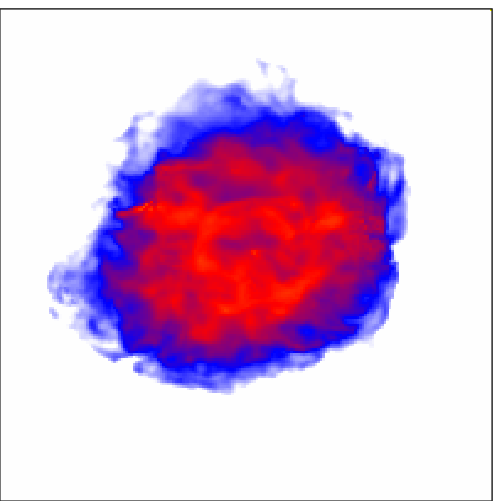}}
\put(0.0,2.5){\includegraphics[width=5.6cm,height=5.6cm]{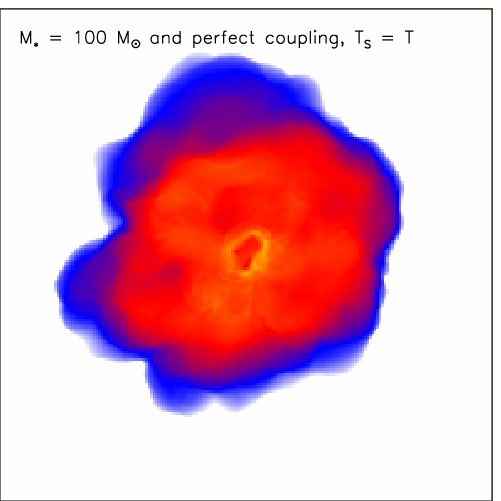}}
\put(5.7,2.5){\includegraphics[width=5.6cm,height=5.6cm]{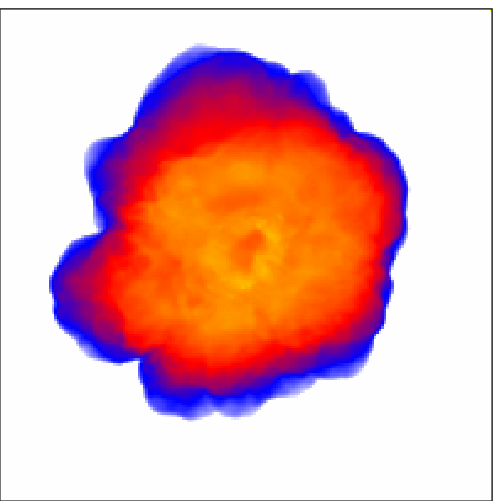}}
\put(11.4,2.5){\includegraphics[width=5.6cm,height=5.6cm]{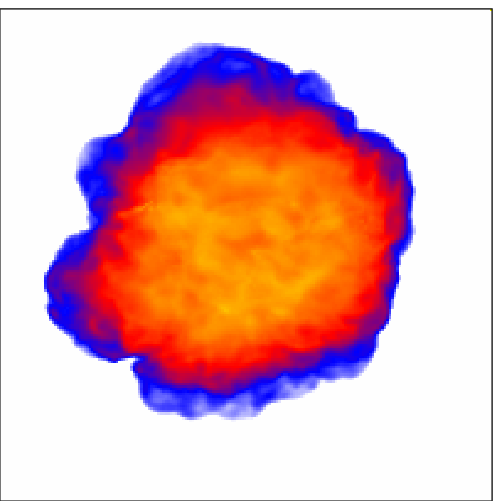}}
\put(0.0,0.0){\includegraphics[width=17cm,height=2.5cm]{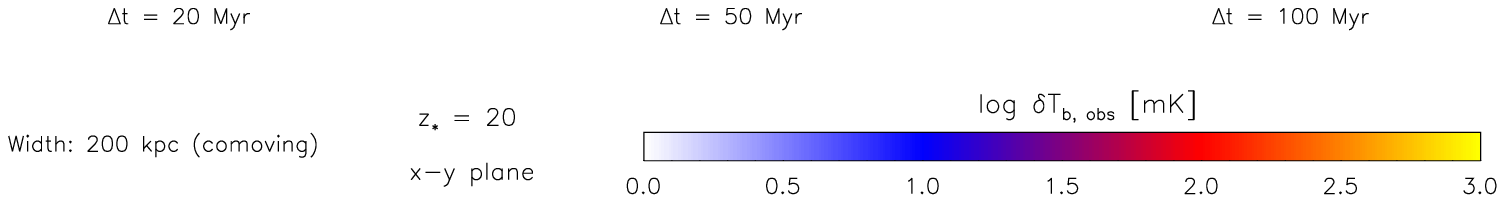}}
\end{picture}}
\caption{The observed differential brightness temperature of the relic H~{\sc ii} region around a $100~{\rm M}_{\odot}$ Pop~III star, shown $20$, $50$ and $100~{\rm Myr}$ after the star has turned off. We delineate the range of possible values by showing the result for collisional coupling only (top row), as well as perfect coupling to the Ly$\alpha$ background, resulting in $y_{\alpha}\gg 1$ or $T_{S}=T$ (bottom row). In the first case, the observed differential brightness temperature is of order a few $10~{\rm mK}$ for $\simeq 100~{\rm Myr}$, while in the second case the signal is of order a few $100~{\rm mK}$ for well over $\simeq 100~{\rm Myr}$. We note that the latter is likely more relevant at $z\la 20$, where the observationally accessible signal is produced \citep[see][]{furlanetto06,pf07}.}
\end{center}
\end{figure*}

The differential brightness temperature with respect to the CMB may then be derived as follows. In the Rayleigh-Jeans limit and for $T_{S}\gg T_{*}$, the monochromatic radiative transfer equation for a ray passing through a cloud, evaluated in its comoving frame, may be written in terms of the brightness temperature $T_{b}$:
\begin{equation}
T_{b}=T_{\gamma}\,e^{-\tau}+\int_{0}^{\tau}T_{S}\,e^{-\tau'}{\rm d}\tau'\mbox{\ ,}
\end{equation}
where the optical depth at the 21-cm line is given by
\begin{equation}
{\rm d}\tau=\frac{3c^{2}A_{21}n_{\rm HI}}{32\pi\nu_{21}^{2}}\,\phi(\nu_{21})\,\frac{T_{*}}{T_{S}}\,{\rm d}s\mbox{\ .}
\end{equation}
Here, $\phi(\nu_{21})$ is the normalized line profile at the resonance frequency $\nu_{21}$ and ${\rm d}s$ is the distance traveled by the ray. In our case, the line profile is dominated by thermal broadening, with a Doppler width given by
\begin{equation}
\Delta\nu_{\rm D}=\nu_{21}\sqrt{\frac{2k_{\rm B}T}{\mu m_{\rm H}c^{2}}}\mbox{\ .}
\end{equation}
The amplitude of the line profile at the resonance frequency may be replaced by the Doppler width, i.e. $\phi(\nu_{21})=\Delta\nu_{\rm D}^{-1}$. With this definition, equation~(35) yields the differential brightness temperature $\delta T_{b}=T_{b}-T_{\gamma}$, which becomes particularly simple for a constant spin temperature and the fact that the relic H~{\sc ii} regions considered here are optically thin:
\begin{equation}
\delta T_{b}=(T_{S}-T_{\gamma})\,\tau\mbox{\ .}
\end{equation}
The observed differential brightness temperature is then simply given by $\delta T_{{\rm b,obs}}=\delta T_{b}/\left(1+z\right)$.

In Fig.~6, we show the observed differential brightness temperature for a $100~{\rm M}_{\odot}$ star and the two limiting cases discussed above. Note that we have only taken into account ionized gas along the line of sight. For collisional coupling, the observed differential brightness temperature is of order a few $10~{\rm mK}$ for $\simeq 100~{\rm Myr}$, while for perfect coupling the signal is elevated by an order of magnitude to a few $100~{\rm mK}$ for well over $\simeq 100~{\rm Myr}$. In reality, the expected signal lies between these extremes and is a function of redshift, since collisional coupling becomes weaker as the background density drops, while Ly$\alpha$ coupling becomes stronger as the Ly$\alpha$ background rises. At $z\la 20$, where the observationally accessible signal is produced, the Ly$\alpha$ background is likely strong enough for the latter to be more important \citep{furlanetto06,pf07}.

Next, we determine the radio background produced by the integrated 21-cm emission of relic H~{\sc ii} regions. The differential specific flux observed at the redshifted 21-cm line from a single relic H~{\sc ii} region with differential brightness temperature $\delta T_{b}$ is given by
\begin{equation}
\delta F_{\nu}=\frac{2k_{\rm B}\nu_{21}^{2}}{c^{2}}\,\left(1+z\right)^{-3}\Delta\Omega\,\delta T_{b}\mbox{\ ,}
\end{equation}
where $\Delta\Omega=A/D_{\rm A}^{2}$ denotes the solid angle subtended by the relic H~{\sc ii} region, $A=\pi r_{\rm HII}^{2}$ its area, $r_{\rm HII}=\left(3N_{\rm ion}/4\pi n_{\rm H,b}\right)^{1/3}$ its radius, and $D_{\rm A}$ the angular diameter distance. The average differential specific flux $\left<\delta F_{\nu}\right>$ within a beam size $\Delta\Omega_{\rm beam}$ and bandwidth $\Delta\nu_{\rm obs}$ is then given by
\begin{equation}
\left<\delta F_{\nu}\right>=\delta F\,N_{\rm ps}(z)\,\frac{{\rm d}^{2}V(z)}{{\rm d}z\,{\rm d}\Omega}\,\frac{\Delta z\,\Delta\Omega_{\rm beam}}{\Delta\nu_{\rm obs}}\mbox{\ ,}
\end{equation}
where $\delta F=\delta F_{\nu}\,\Delta\nu_{\rm D}/\left(1+z\right)$, $N_{\rm ps}(z)$ is the Press-Schechter mass function defined in equation~(24), $\Delta z=\Delta\nu_{\rm obs}\left(1+z\right)^{2}/\nu_{21}$, and ${\rm d}^{2}V(z)/{\rm d}z\,{\rm d}\Omega$ is the comoving volume per unit redshift and solid angle:
\begin{equation}
\frac{{\rm d}^{2}V(z)}{{\rm d}z\,{\rm d}\Omega}=\frac{c\,D_{\rm A}^{2}\left(1+z\right)^{2}}{H(z)}\mbox{\ ,}
\end{equation}
where $H(z)$ is the Hubble expansion rate. With the definition of the brightness temperature, the average differential antenna temperature $\left<\delta T_{b}\right>$ is given by
\begin{equation}
\left<\delta T_{b}\right>=\frac{\pi c}{\nu_{21}}\,\frac{\left(1+z\right)^{2}N_{\rm ps}(z)}{H(z)}\,\Delta\nu_{\rm D}\,r_{\rm HII}^{2}(z)\,\delta T_{b}(z)\mbox{\ .}
\end{equation}
Based on our argument above, we assume that the Ly$\alpha$ background is strong enough for perfect coupling at all redshifts. In this case, and for $T\gg T_{\gamma}$, the average differential antenna temperature becomes independent of electron fraction and temperature:
\begin{equation}
\left<\delta T_{b}\right>=\frac{9c^{3}A_{21}T_{*}N_{\rm ion}}{128\pi \nu_{21}^{3}H_{0}\sqrt{\Omega_{m}}}\,\left(1+z\right)^{1/2}N_{\rm ps}(z)\mbox{\ ,}
\end{equation}
where we have set $n_{\rm HI}=n_{\rm H,b}$ in equation~(36). We note that the observed frequency is related to the redshift via $\nu_{\rm obs}=\nu_{21}/\left(1+z\right)$. We have further assumed that the relic H~{\sc ii} region produced by each star-forming minihalo persists until the Universe is reionized (i.e. $z\simeq 6$), which is a good approximation for perfect coupling and $T\gg T_{\gamma}$. Equation~(43) thus provides a robust upper limit for the collective 21-cm emission from the first relic H~{\sc ii} regions.

In Fig.~7, we compare the average differential antenna temperature and specific flux for a beam size of $\Delta\theta_{\rm beam}=15'$ to the sensitivity of the SKA, assuming a $10\,\sigma$ detection, a bandwidth of $\Delta\nu_{\rm obs}=1~{\rm MHz}$, and an integration time of $1000~{\rm h}$. At all frequencies, the maximum 21-cm signal from the first relic H~{\sc ii} regions is of order $10~{\rm mK}$, which is well detectable by the SKA. The effects of biasing and radiative feedback will reduce this signal, but probably not enough to fall below the sensitivity of the SKA. Compared to free-free emission, the 21-cm signal is typically an order of magnitude stronger, and offers the best prospect for indirectly probing the first stars. Furthermore, the 21-cm signal is explicitly frequency-dependent, while this is not the case for bremsstrahlung, where a flat spectrum is produced (see equation~28). This dependency might allow for a better distinction from other sources of radio emission at these wavelengths.

\begin{figure}
\begin{center}
\includegraphics[width=8.5cm]{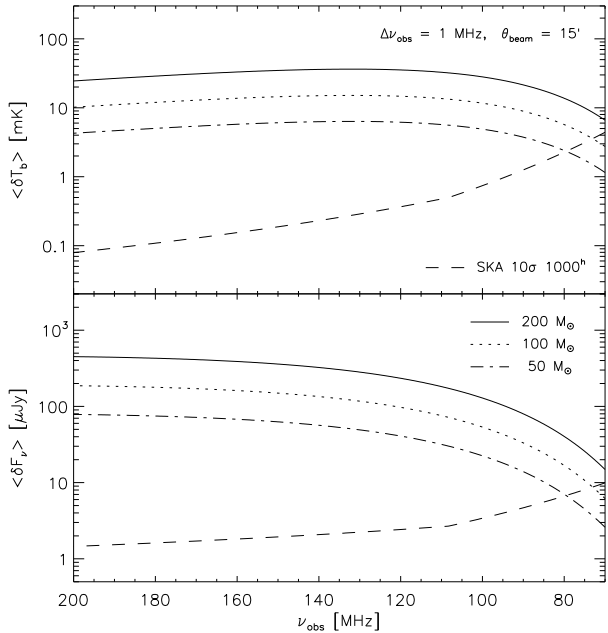}
\caption{The average differential antenna temperature and specific flux of the radio background produced by 21-cm emission, shown as a function of observed frequency. We have chosen a beam size of $15~{\rm arcmin}$ to achieve the highest possible resolution and sensitivity at $100~{\rm MHz}$ for the currently planned configuration of the SKA. The dot-dashed, dotted and solid lines correspond to an initial mass function consisting solely of $50$, $100$ and $200~{\rm M}_{\odot}$ Pop~III stars, respectively. The dashed line shows the sensitivity of the SKA for a $10\,\sigma$ detection, a bandwidth of $1~{\rm MHz}$, and an integration time of $1000~{\rm h}$. In all cases, the 21-cm signal is well above the detection threshold of the SKA. The effects of biasing and radiative feedback will reduce this signal, but probably not enough to fall below the sensitivity of the SKA. Compared to free-free emission, the 21-cm signal is typically an order of magnitude stronger, and offers the best prospect for indirectly probing the first stars.}
\end{center}
\end{figure}

\section{Summary and conclusions}
We have introduced a general-purpose radiative transfer scheme for cosmological SPH simulations that treats ionizing and photodissociating radiation from massive Pop~III stars in the early Universe. Based on this methodology, we have investigated the build-up of the first H~{\sc ii} regions and relic H~{\sc ii} regions around Pop~III stars formed in minihaloes, and predicted their contribution to the extragalactic radio background via bremsstrahlung and 21-cm emission. Although recombination radiation from individual H~{\sc ii} regions in their active as well as relic states is too faint to be directly detectable even with {\it JWST}, their collective radio emission might be strong enough to be within reach of the planned SKA. In particular, we have found that the integrated free-free emission results in a maximum differential antenna temperature of $\simeq 1~{\rm mK}$, while the 21-cm emission is an order of magnitude stronger. Considering the effects of biasing and negative radiative feedback, which would act to reduce the predicted signal, the free-free signal is likely beyond the capability of the SKA, while the 21-cm signal will most likely be observable, providing an excellent opportunity for indirectly probing the first stars.

We note that an analysis of the angular fluctuation power spectrum will be essential to isolate the 21-cm signal from other backgrounds \citep{fo06}, although the frequency-dependence of the 21-cm signal might already prove useful. Among these are neutral minihaloes, which appear in emission due to their enhanced density and temperature \citep{iliev02}, or IGM gas heated by X-rays from supernovae \citep{oh01}, X-ray binaries \citep{gb03}, or the first quasars \citep{madau04b,kmm06}. A strong absorption signal might originate from cold, neutral gas if the Ly$\alpha$ background effectively couples the spin temperature to the gas temperature \citep{pf07}. In addition, there is the signal produced by stars (primordial or already metal-enriched) formed in the first dwarf galaxies \citep[e.g.][]{nb08}. All of these compete with each other, and more work is required to understand their relative importance. One important task is to extend the simulations to larger cosmological volumes, to measure the aggregate signal from many sources in a more robust way.

Minihaloes may not have been the dominant formation sites for primordial stars, in terms of producing the bulk of the radiation that drove reionization, or of being the source for the majority of the heavy elements present at high redshifts \citep{gb06,sbk08}. Nevertheless, they are the ideal laboratory to test our current standard model of the first stars, by providing an exceedingly simple environment for the star formation process \citep{bromm09}. The next step in the hierachical build-up of structure is already highly complex, due to the presence of metals, turbulent velocity fields, and possibly dynamically significant magnetic fields \citep{wa07b,wa08b,greif08,schleicher09}. It is therefore crucial to empirically probe the minihalo environment, and the signature left in the radio background might provide us with one of the few avenues to accomplish this in the foreseeable future.

\section*{Acknowledgments}
The authors would like to thank the referee Naoki Yoshida for his valuable comments and suggestions that greatly improved this work. TG thanks Matthias Bartelmann and Simon Glover for many stimulating discussions. TG acknowledges financial support by the Heidelberg Graduate School of Fundamental Physics (HGSFP). The HGSFP is funded by the Excellence Initiative of the German government (grant number GSC 129/1). RSK acknowledge subsidies from the {\em Deutsche Forschungsgemeinschaft} via the priority program SFB 439 ``Galaxies in the Early Universe'' as well as via grants KL1358/1, KL1358/4 and KL1358/5. In addition, RSK also acknowledges partial support from a  Frontier grant of Heidelberg University funded by the German Excellence Initiative. VB acknowledges support from NSF grant AST-0708795 and NASA ATFP grant NNX08AL43G. The simulations presented here were carried out at the Texas Advanced Computing Center (TACC).

\bibliographystyle{mn2e}

\end{document}